\documentclass[review]{elsarticle}

\usepackage{lineno,hyperref}
\usepackage{graphicx}
\usepackage{amsmath}
\usepackage{amssymb}
\usepackage{physics}
\modulolinenumbers[5]

\journal{Nuclear Physics A}

\begin{document}

\begin{frontmatter}

\title{Coulomb field correction due to virtual $e^+ e^-$ production in heavy ion collisions}

\author[physicsdept,cyclotron]{Thomas Settlemyre}

\author[shaanxi]{Hua Zheng}

\author[cyclotron,catania]{Aldo Bonasera}
\address[physicsdept]{Department of Physics and Astronomy, Texas A\&M University, College Station, TX 77843, United States of America}

\address[cyclotron]{Cyclotron Institute, Texas A\&M University, College Station, TX 77843, United States of America}

\address[shaanxi]{School of Physics and Information Technology, Shaanxi Normal University, Xi’an 710119, People’s Republic of China}

\address[catania]{Laboratori Nazionali del Sud, INFN, via Santa Sofia, 62, 95123 Catania, Italy}

\begin{abstract}
The correction to the Coulomb energy due to virtual production of $e^+e^-$ pairs, which is on the order of one percent of the Coulomb energy at nuclear scales is discussed. The effects of including a pair-production term in the semi-empirical mass formula and the correction to the Coulomb barrier for a handful of nuclear collisions using the Bass and Coulomb potentials are studied. With an eye toward future work using Constrained Molecular Dynamics (CoMD) model, we also calculate the correction to the Coulomb energy and force between protons after folding with a Gaussian spatial distribution.
\end{abstract}


\end{frontmatter}

\linenumbers

\section{Introduction}

The Coulomb force is mediated by the exchange of a virtual photon. It is possible for a virtual electron-positron pair to be created and annihilated in this process. These virtual charges tend to polarize the vacuum, resulting in a correction to the $1/r$ potential. In 1935, Uehling derived the vacuum polarization correction to first order in the fine structure constant $\alpha$ \cite{uehling}. This correction is important in the analysis of $p$-$p$ scattering data \cite{preston}. We will show in this paper that the vacuum polarization correction is on the order of one percent of the Coulomb energy in nuclear collision systems. One percent may seem small, but the strong fields in fission processes can be on the order of 200 MeV. A correction on the order of 2 MeV could noticeably affect the height of the Coulomb barrier, where the nuclear and Coulomb energies roughly cancel, thus the outcome of the sub-barrier fusion. A larger Coulomb barrier increases the cross-section of sub-barrier fusion, e.g. Carbon-Carbon fusion in the cores of stars. 
Like the regular Coulomb energy, the vacuum polarization correction is proportional to $Z^2$. Thus, the correction is larger for proton-rich nuclei. The resulting change in binding energy could affect the proton dripline. 

The vacuum polarization is not just a perturbative effect; production of real $e^+e^-$ pairs can occur during dynamics in the presence of strong fields, when the available energy exceeds twice the electron mass \cite{schwinger,wong,blaschke}. In the 1980s, experimentalists at GSI found some anomalous production of $e^+e^-$ pairs in heavy ion collisions.
Various explanations were proposed, including production of a hypothesized new light particle and experimental error \cite{Schafer_1989}. To our knowledge, there is no consensus \cite{Griffin1998}.

In this first paper we only discuss the perturbative effect on the energy. We will introduce the correction into the microscopic model Constrained Molecular Dynamics (CoMD) \cite{comd1,comd2,comd3,comd4,comd5,comd6,comd7}, and in following research we will discuss actual production.
The structure of this paper is as follows. In section \ref{section:uehling}, we summarize the result of Uehling \cite{uehling} for the vacuum polarization correction to the Coulomb potential. In section \ref{section:mass} we derive a modified semi-empirical mass formula that includes a term for the vacuum polarization energy. Section \ref{section:bass} includes the Bass potential to model the interaction of two nuclei. In section 5, the vacuum polarization correction is compared to the total potential energy. We conclude with a calculation of the form of the proton-proton energy after folding with a Gaussian distribution to be used in future work with Constrained Molecular Dynamics (CoMD) \cite{comd2} and/or other models \cite{aichelin}. A brief summary is given in section 6.

\section{General formulas}
\label{section:uehling}

\begin{figure}
    \centering
    \includegraphics[width=.3\textwidth]{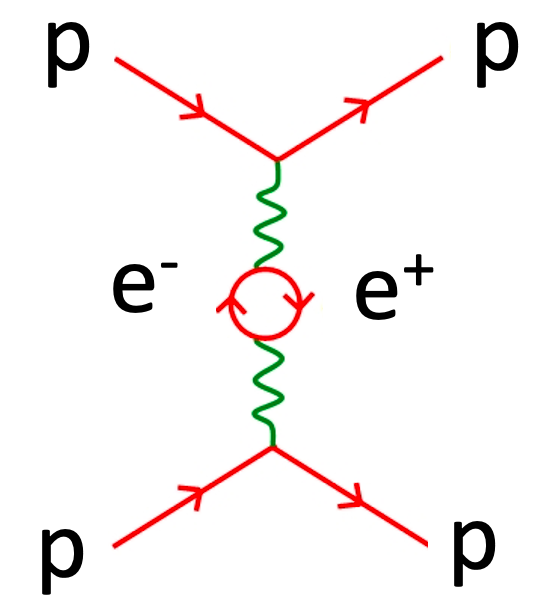}
    \caption{The Feynman diagram for the process where a  virtual $e^+e^-$ pair is created and annihilated in the exchange of a photon between protons.}
    \label{fig:feynman}
\end{figure}

The usual inverse-square Coulomb force between charged particles comes from the exchange of a virtual photon. It is possible that a virtual electron-positron pair is created and annihilated during the exchange. (The Feynman diagram for this process is shown in Figure \ref{fig:feynman}.) The presence of virtual charged particles creates a polarization in the vacuum, which gives a correction to the Coulomb energy.

Uehling \cite{uehling} showed that the correction to the Coulomb energy between two nuclei of charge $Z_1 e$ and $Z_2 e$ separated by a distance $r$ is
\begin{equation}
    V_{e^+e^-}(r) = - \frac{\alpha Z_1 Z_2 e^2}{\pi r} \int_0^1(1-u^2) li\left( \exp{\frac{-2r}{\lambda_0 \sqrt{1-u^2}}}\right) du, \label{eq:complicatedpot}
\end{equation}
where $\lambda_0 = \hbar/m_e c \approx 386 \, \text{fm}$ is the (reduced) Compton wavelength of the electron, and
\begin{equation}
    li(x) = \int_0^{x} \frac{dt}{\log{t}},
\end{equation}
is the logarithmic integral function.

\begin{figure}
    \centering
    \includegraphics[width=.6\textwidth]{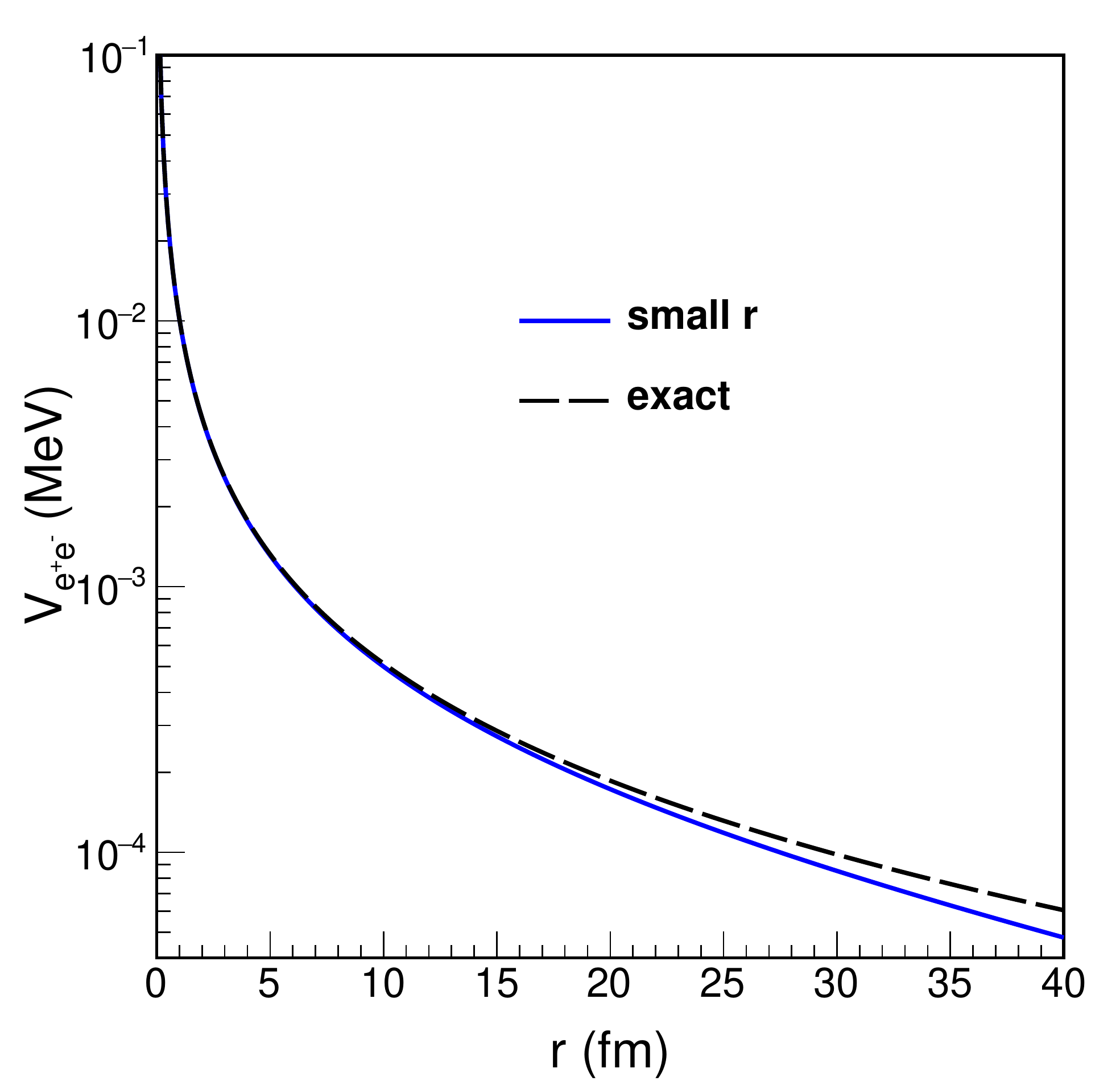}
    \caption{(Color online) The comparison between the small-$r$ approximation (Eq. \eqref{eq:uehlingpot}, the solid blue line) and 
    the exact expression (Eq. \eqref{eq:complicatedpot}, the dashed black line) for the correction to Coulomb energy in p-p scattering.}
    \label{fig:smallexact}
\end{figure}

For $r \ll \lambda_0$, Eq. \eqref{eq:complicatedpot} is approximated by the simpler expression \cite{uehling}
\begin{equation}
   V_{e^+e^-}(r) =  -\frac{2\alpha}{3\pi}\frac{Z_1 Z_2 e^2}{r}\left( \ln\frac{r}{\lambda_0} + \gamma + \frac{5}{6} \right), \label{eq:uehlingpot}
\end{equation}
where $\gamma = 0.5772...$ is the Euler-Mascheroni constant. We will use exclusively the simpler result from Eq. \eqref{eq:uehlingpot}, which is a very good approximation on nuclear scales (see Fig. \ref{fig:smallexact}).


\section{Vacuum polarization term in mass formula}
\label{section:mass}

\begin{figure}
    \centering
    \includegraphics[width=.75\textwidth]{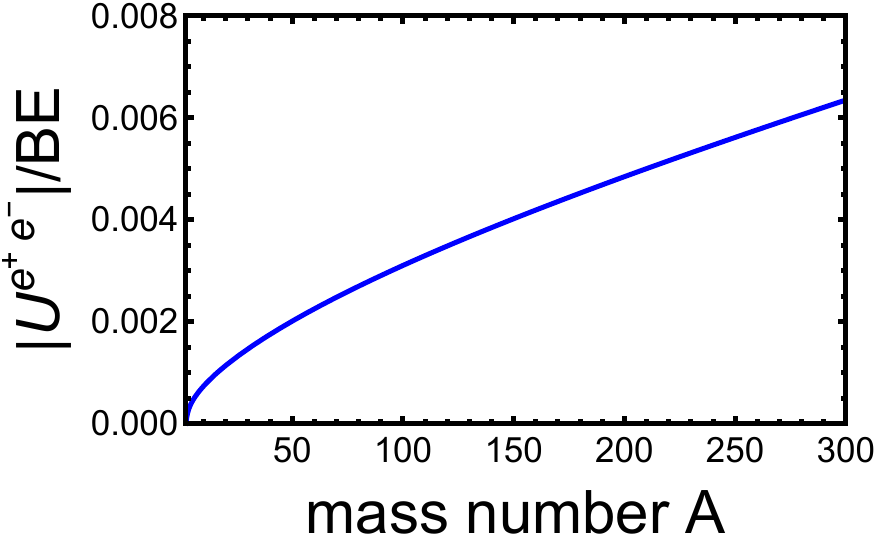}
    \caption{The relative change to binding energy when the vacuum polarization term is included. $Z$ is chosen to minimize the binding energy according to Equation \eqref{eq:zmin}. The correction decreases the binding energy.  The change is on the order of half a percent, but has more of an effect for larger $A$.}
    \label{fig:isobarplot}
\end{figure}

\begin{figure}
    \centering
    \includegraphics[width=.75\textwidth]{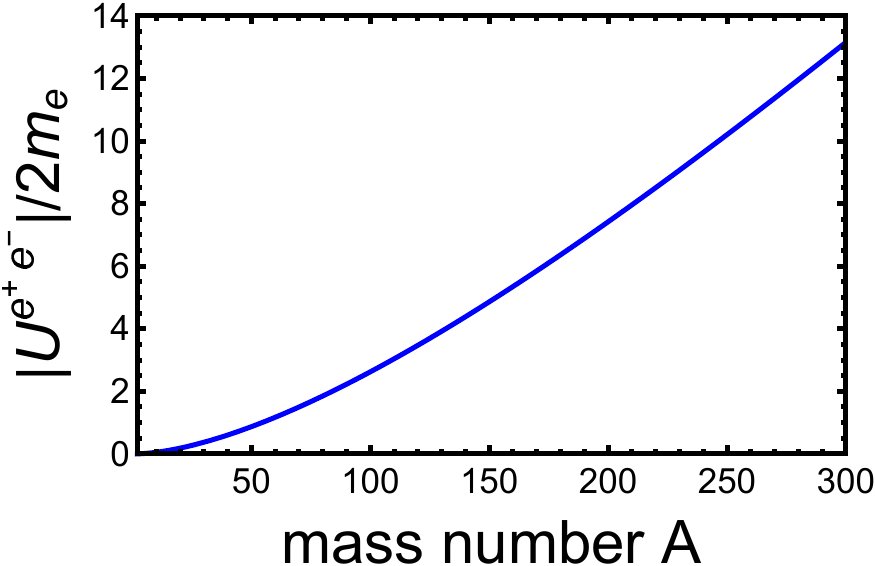}
    \caption{The $e^+e^-$ correction in units of the mass of an $e^+e^-$ pair. $Z$ is chosen to minimize the binding energy according to Equation \eqref{eq:zmin}. The correction is greater than $2m_e$ in absolute value for $A \geq 55$.}
    \label{fig:correctionplot}
\end{figure}

The binding energy of a nucleus according to the liquid drop model is \cite{krane}, \cite{preston}
\begin{equation}
    BE = a_V A - a_s A^{2/3} - a_C Z(Z-1) A^{-1/3} - a_{sym} \frac{(A-2Z)^2}{A} + \delta. \label{eq:massformula}
\end{equation}
In particular, the term $a_C Z(Z-1) A^{-1/3}$ is the Coulomb self-energy of the charge in the nucleus. If we consider the correction to the Coulomb energy due to vacuum polarization, there will be an analogous correction term in the mass formula (Eq. \eqref{eq:massformula}). To derive the form of this correction, we assume the nucleus has charge of $Ze$ uniformly distributed inside a sphere of radius $R = 1.2 A^{1/3}$ fm. The charge density is
\begin{equation}
    \rho = \frac{Ze}{4\pi R^3/3}.
\end{equation}
The correction to the Coulomb self energy inside the nucleus is
\begin{align}
    U_{e^+e^-} &= \frac{1}{2}\int \rho d^3r \int \rho d^3r' \left[ -\frac{2\alpha}{3\pi}\frac{e^2}{|\mathbf{r}-    \mathbf{r}'|}\left( \ln\frac{|\mathbf{r} -    \mathbf{r}'|}{\lambda_0} + \gamma + \frac{5}{6}\right) \right] \nonumber \\
    &= -\frac{2\alpha}{3\pi} \left(\gamma + \frac{5}{6} \right) U_{Coul} - \frac{2\alpha}{3\pi} U_{Log},
\end{align}
where $U_{Coul}$ is the standard Coulomb self-energy and
\begin{equation}
    U_{Log} = \int \rho d^3r \int \rho d^3r' \frac{\ln{\left(|\mathbf{r} -    \mathbf{r}'|/\lambda_0\right)}}{|\mathbf{r} -    \mathbf{r}'|}.
\end{equation}
The integral evaluates exactly to
\begin{align}
    U_{Log} &=-\frac{31Z^2e^2}{50R} + \frac{3Z^2e^2}{5R} \ln{\frac{2R}{\lambda_0}} \nonumber \\
    &= \frac{3e^2}{5r_0} \, \frac{Z^2}{A^{1/3}} \left(-\frac{31}{30} + \ln{\left(\frac{2r_0}{\lambda_0}\right)} + \frac{1}{3}\ln{A}\right).
\end{align}
We identify the factor in front as the Coulomb energy term in the mass formula by making the standard replacement $Z^2 \rightarrow Z(Z-1)$ so that the Coulomb self-energy of the proton is zero. The correction term is thus
\begin{align}
    U_{e^+e^-} &= -\frac{2\alpha}{3\pi}\, a_C \frac{Z(Z-1)}{A^{1/3}}\left(\gamma - \frac{1}{5} + \ln{\left(\frac{2r_0}{\lambda_0}\right)} + \frac{1}{3}\ln{A} \right) \nonumber \\
    &= (0.0073 - 0.00052 \ln{A}) a_C \frac{Z(Z-1)}{A^{1/3}}.
\end{align}
The $e^+e^-$ term decreases the binding energy. Figure \ref{fig:isobarplot} shows the binding energy with and without the correction for the isobars that minimize the binding energy, \cite{krane}
\begin{equation}
    Z_{min} = \frac{A}{2} \left( \frac{1}{1+\frac{1}{4}A^{2/3} a_c/a_{sym}} \right) \label{eq:zmin}.
\end{equation}
The difference is small, but becomes larger as $A$ increases. When $A \geq 55$, the correction becomes larger in magnitude than $2m_e$, the mass of an $e^+e^-$ pair (Figure \ref{fig:correctionplot}). This indicates the size of nuclei where one would expect production of one or more real $e^+e^-$ pairs in collisions.

\section{Coulomb barrier in nucleus-nucleus collisions}
\label{section:bass}
\begin{figure}
    \centering
    \includegraphics[width=.6\textwidth]{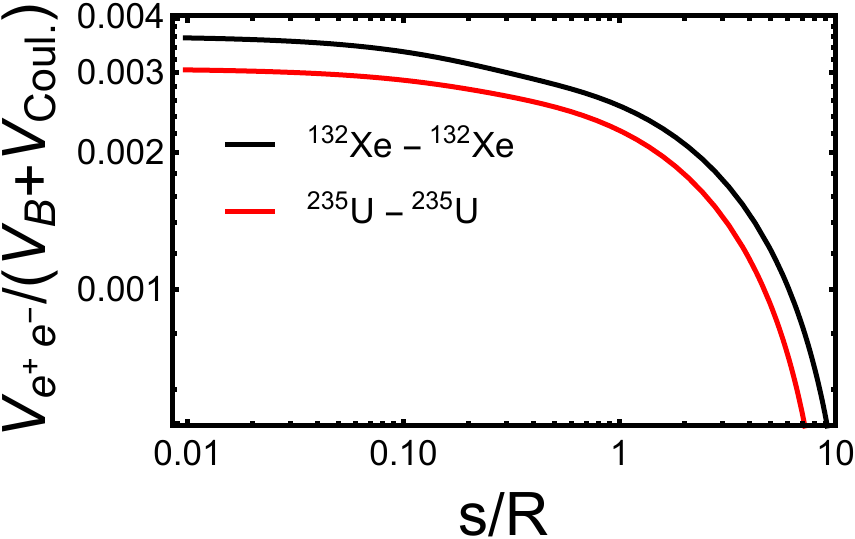}
    \caption{Relative correction to the nuclear potential as a function of $s$ for the collisions indicated. The correction is on the order of a third of a percent when the nuclei are close. Distances are in units of the nuclear radius.}
    \label{fig:relbass}
\end{figure}

Until now, we have only discussed the Coulomb potential and a correction to it from vacuum polarization. Bass \cite{bass} derived a phenomenological nuclear potential from experimental fusion cross sections assuming negligible friction at large distances. For each nucleus involved the radius is estimated as
\begin{equation}
    R = a A^{1/3} - b^2 (a A^{1/3})^{-1},
\end{equation}
with $a = 1.16 \, \text{fm}$, $b^2/a = 1.39\,\text{fm}$. The Bass potential is, in terms of the distance between the nuclear surface $s = r - R_1 - R_2$,
\begin{equation}
    V_B(s) = - \frac{R_1 R_2}{R_1 + R_2} [ \alpha \exp(s/d_1) + \beta \exp(s/d_2) ]^{-1},
\end{equation}
with $\alpha = 0.0300\, \text{MeV}^{-1}\, \text{fm}$, $\beta = 0.0061\, \text{MeV}^{-1}\, \text{fm}$, $d_1 = 3.30\,\text{fm}$, and $d_2=0.65\,\text{fm}$, which are obtained by fitting scattering data. It gives the potential energy between nuclei as a function of the distance $s$ between their surfaces. In Figure \ref{fig:relbass} we can see that the vacuum polarization energy is on the order of a third of a percent to a tenth of a percent of the total potential energy (Bass + Coulomb) for $s$ less than 5 times the nuclear radius for $\null^{132}$Xe - $\null^{132}$Xe and $\null^{235}$U - $\null^{235}$U collisions.

\subsection{Sub-barrier fusion}

We have shown in the previous sections that the $e^+e^-$ correction to the Coulomb field is rather modest, less than 1\%.  The correction may become more important in the fusion of two heavy ions below the Coulomb energy and especially in the Gamow region relevant for nuclear Astrophysics process and stellar evolution \cite{x1}.  In a recent paper \cite{x2}, a macroscopic model dubbed ‘neck model’ \cite{x3} has been extended to collisions below the Coulomb barrier using the Feynman path integral method \cite{x4} i.e. extending the dynamics below the barrier to imaginary times. Recently, the $\null^{12}$C+$\null^{12}$C reaction at Gamow energies has received a great attention both theoretically \cite{x2,x5,x6,x7,x8} and experimentally \cite{x9,x10,x11,x12,x13,x14,x15} because of its relevance in carbon burning stars. Thus it may be of importance to estimate the effect of $e^+e^-$ virtual pairs in these reactions. In the model, the two nuclei approach each other under the action of the Coulomb plus the nuclear (Bass) potential. For beam energies below the Coulomb barrier, the two ions classically stop at the point of closest approach and bounce back. Here, we will consider only central collisions i.e. zero impact parameter. Quantum mechanically there is a small and finite probability of tunneling the barrier. Using the Feynman path method, we solve the classical equation of motion in imaginary time starting at the point of closest distance or outer external point \cite{x2,x4}. The action is calculated during the imaginary time propagation up to the second or inner turning point. At the inner turning point we switch back to real time but now the nuclei are well inside the nuclear field and they fuse. From the action we can calculate the probability of tunneling and the fusion cross section.  Since such quantity is extremely small, it is customary to express it in terms of the astrophysical S-factor (or the modified S* factor), which takes into account the Coulomb penetration factor \cite{x1}:
\begin{align}
    S^*(E_{c.m.})&=E_{c.m.}\sigma(E_{c.m.}) \times \exp(87.12E_{c.m.}^{-1/2}+0.46Ec.m.) \nonumber \\
    &=S(E_{c.m.})  \times \exp(0.46E_{c.m.}). \label{eq:x1}
\end{align}

\begin{figure}
    \centering
    \includegraphics[width=.6\linewidth]{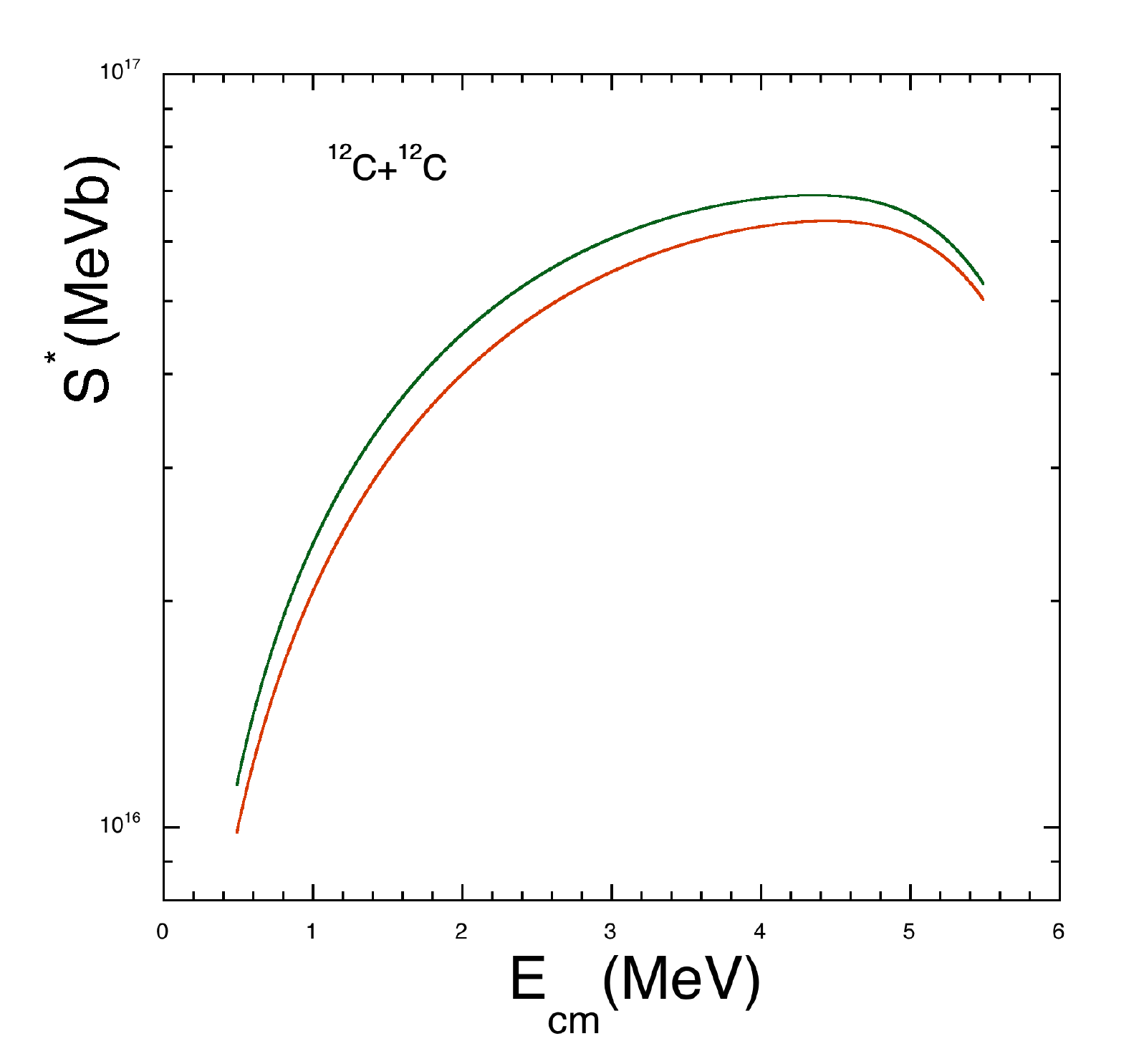}
    \caption{Modified astrophysical S*-factor as function of the center of mass energy for the $\null^{12}$C+$\null^{12}$C system.  The lower curve is obtained adding the $e^+e^-$ correction, Eq.\eqref{eq:uehlingpot}, to the Coulomb interaction.}
    \label{fig:x1}
\end{figure}

In figure \ref{fig:x1}, we plot the modified S* factor as function of the center of mass energy with (lower curve) and without (upper curve) $e^+e^-$ correction to the Coulomb field. As expected the modification is quite relevant in this energy region reaching about 15\% at the lowest energies.

\subsection{Fission}

The $e^+e^-$ correction to the Coulomb field may be relevant in the dynamics of fission. In order to estimate its influence on the final kinetic energy of the fission fragments we can use the ‘neck model’ \cite{x2,x3} and in particular its application to symmetric fission \cite{x16}. In the model, the fission dynamics approximately from the saddle point to the fission point is geometrically described as two half-spheres of radius $R$ (the final fragments) joined by two sections of cones of radii $R$ and $r_N$ (the neck radius) respectively \cite{x3}.  Assuming incompressible matter, the total volume during fission is equal to the volume of the two separated fragments. This gives the following relation between $r_N$ and the relative distance $r$:
\begin{equation}
    r_N=\frac{\sqrt{r^2 R^2-4r(rR^2-4R^3)-Rr}}{2r}.	\label{eq:x2}
\end{equation}
Fission occurs when the system becomes very elongated as compared to the neck radius \cite{x17}.  This leads to the so-called Rayleigh instability and we assume to occur when $r_N<1\,\text{fm}$.  At this stage, the final kinetic energy of the fission fragments is given by the difference between the Coulomb repulsion and the nuclear attraction \cite{x16}.  Here we are not interested in the detailed dynamics of fission but just an estimate of the $e^+e^-$ contribution at the fission point.  The relative distance $r$ between the fragments at fission maybe estimated from Eq.\eqref{eq:x2} assuming $r_N=1\,\text{fm}$ and solving for $r$.

\begin{figure}
    \centering
    \includegraphics[width=.75\linewidth]{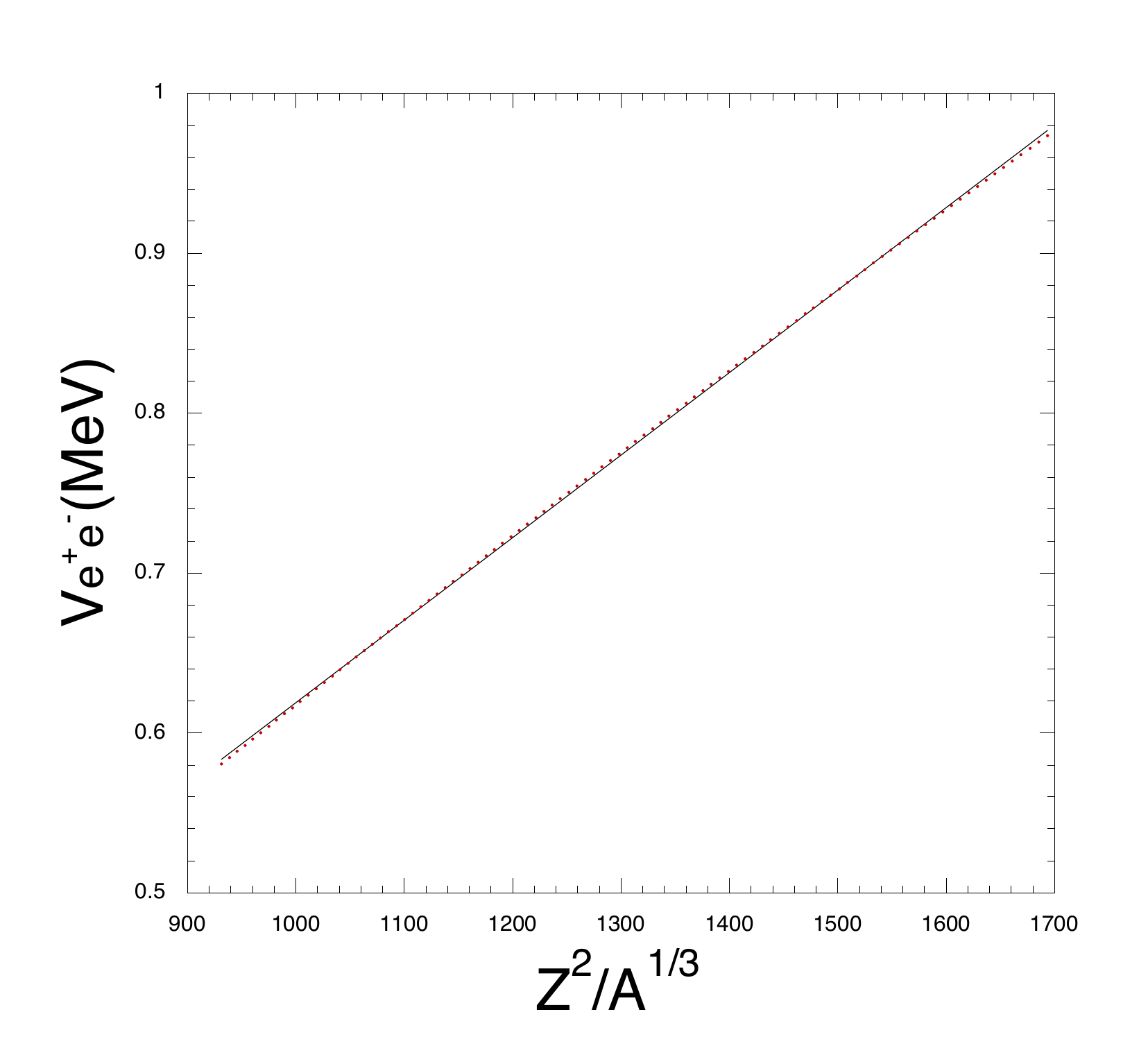}
    \caption{$e^+e^-$ Correction as function of $Z^2/A^{1/3}$ \cite{x18}  }
    \label{fig:x2}
\end{figure}

Using Eq.\eqref{eq:zmin} we obtain the most probable $Z_{min}$ for each $A$, the fragment radii $R$ and the distance $r$ at the fission point from Eq. \eqref{eq:x2}. Inserting $r$ into Eq.\eqref{eq:uehlingpot}, we can calculate $V_{e^+e^-}$ and plot it in figure \ref{fig:x2} as function of $Z^2/A^{1/3}$.  The correction is of the order of 0.5 MeV for parent nuclei $A=180$ to reach 1.0 MeV for $A=280$.  The plot displays a linear dependence fit as:
\begin{equation}
V_{e^+ e^-}=0.0005 Z^2/A^{1/3} +0.1\, \text{MeV}. \label{eq:x3}
\end{equation}
This can be compared to the Viola systematics \cite{x18} for the final kinetic energy of the fission fragments $\langle E_k \rangle$ with the parameters fitted to the experimental data for symmetric fission:
\begin{equation}
\langle E_k \rangle =0.1189 Z^2/A^{1/3} +7.3 \,\text{MeV}. \label{eq:x4}
\end{equation}
A comparison between Eq.\eqref{eq:x3} and Eq. \eqref{eq:x4} reveals a contribution of the Coulomb correction less than 1\% as expected.

\section{Folding for CoMD}
\label{section:comd}
When considering the interaction of nucleons at short enough distances, the point particle assumption is not proper and the size of the nucleon should be taken into account. In CoMD \cite{comd1,comd2,comd3,comd4,comd5,comd6,comd7}, the nucleons are assumed to have a Gaussian distribution with width $\sigma_r$
\begin{align}
    \rho_i(\mathbf{r}) = \frac{1}{(\sqrt{2\pi} \sigma_r)^3}\exp\left[-\frac{\left(\mathbf{r} - \langle \mathbf{r}_i  \rangle \right)^2 }{2\sigma_r^2}\right].
\end{align}
A typical value for $\sigma_r$ is 1.15 fm. Suppose the energy between two nucleons $i$ and $j$ separated by a distance $a$ would be given by $E(a)$ if they were point particles. The folded energy is
\begin{align}
    V(a) &= \int d^3\mathbf{r}_i d^3\mathbf{r}_j E(|\mathbf{r}_i - \mathbf{r}_j|) \rho_i(\mathbf{r}_i) \rho_j(\mathbf{r}_j).
\end{align}
From here we use the same substitutions as refs. \cite{aichelin} and \cite{comd2}. Namely, $\mathbf{r}_1 = \mathbf{r}_i - \langle \mathbf{r}_i \rangle$, $\mathbf{r}_2 = \mathbf{r}_j - \langle \mathbf{r}_j \rangle$, and $\mathbf{a} = \langle\mathbf{r}_i \rangle - \langle\mathbf{r}_j\rangle$, so
\begin{align}
    V(a) &= \frac{1}{(2\pi\sigma_r^2)^3} \int d^3\mathbf{r}_1 d^3\mathbf{r}_2 E(|\mathbf{r}_1 - \mathbf{r}_2 + \mathbf{a}|) \exp \left[-\frac{\mathbf{r}_1^2 + \mathbf{r}_2^2}{2\sigma_r^2}\right].
\end{align}
We take $\mathbf{R} = \mathbf{r}_1 + \mathbf{r}_2$ and $\mathbf{r} = \mathbf{r}_1 = \mathbf{r}_2$. This substitution gives
\begin{align}
    \mathbf{R}^2 + \mathbf{r}^2 = 2(\mathbf{r}_1^2 + \mathbf{r}_1^2), \\
    d^3\mathbf{R} d^3\mathbf{r} = 2^3 d^3\mathbf{r}_1 d^3\mathbf{r}_2,
\end{align}
so the energy is
\begin{align}
    V(a) &=  \frac{1}{(4\pi\sigma_r^2)^3} \int d^3\mathbf{R} d^3\mathbf{r} E(|\mathbf{r} + \mathbf{a}|) \exp \left[-\frac{\mathbf{R}^2 + \mathbf{r}^2}{4\sigma_r^2}\right] \nonumber \\
    &= \frac{1}{(4\pi\sigma_r^2)^{3/2}} \int d^3\mathbf{r} E(|\mathbf{r} + \mathbf{a}|) \exp \left[-\frac{\mathbf{r}^2}{4\sigma_r^2}\right] \nonumber \\
    &= \frac{2\pi}{(4\pi\sigma_r^2)^{3/2}} \int_0^\infty dr\, r^2 \exp \left[-\frac{r^2}{4\sigma_r^2}\right] \int_{-1}^1 d(\cos{\theta}) E(|\mathbf{r} + \mathbf{a}|). \label{polar}
\end{align}

Since the integral does not depend on the direction of $\mathbf{a}$, we can choose $ \mathbf{a}$ along the $z$ direction. We make another change of variables
\begin{align}
    r' \equiv |\mathbf{r}+\mathbf{a}| = \sqrt{r^2 + a^2 + 2ar\cos{\theta}},
\end{align}
and the angular integral becomes
\begin{align}
    \int_{-1}^1 d(\cos{\theta}) E(|\mathbf{r} + \mathbf{a}|) = \frac{1}{ar} \int_{|r-a|}^{|r+a|} dr' r' E(r').
\end{align}
Note that the limits of the integral are $|r\pm a|$, not $|\mathbf{r} \pm \mathbf{a}|$. The energy is
\begin{align}
    V(a) = \frac{2\pi}{(4\pi\sigma_r^2)^{3/2}} \int_0^\infty dr \frac{r}{a} \exp \left[-\frac{r^2}{4\sigma_r^2}\right] \int_{|r-a|}^{|r+a|} dr' r' E(r'). \label{last}
\end{align}

Equation \eqref{last} is a general formula for folding any potential $E(r)$ into the Gaussian distribution. As a check, we re-derive the result for the Coulomb potential between two protons $E(r) = e^2/r$,
\begin{align}
    V_c(a) &= \frac{2\pi e^2}{(4\pi\sigma_r^2)^{3/2}} \int_0^\infty dr \frac{r}{a} \exp \left[-\frac{r^2}{4\sigma_r^2}\right] (|r+a| -|r-a|) \nonumber \\
    &= \frac{4\pi e^2}{(4\pi\sigma_r^2)^{3/2}}\left( \int_0^a dr \frac{r^2}{a} \exp \left[-\frac{r^2}{4\sigma_r^2}\right]  + \int_a^\infty dr\, r \exp \left[-\frac{r^2}{4\sigma_r^2}\right] \right) \nonumber \\
    &= \frac{e^2}{a}\text{erf}\left(\frac{a}{2\sigma_r}\right).
\end{align}
Somewhat more generally, if the energy is given by a power law (PL), $E(r)=r^s$, then
\begin{align}
    V_{PL}(a) = \frac{2\pi}{(4\pi\sigma_r^2)^{3/2}} \int_0^\infty dr \frac{r}{a} \exp \left[-\frac{r^2}{4\sigma_r^2}\right] \frac{|r+a|^{s+2} - |r-a|^{s+2}}{s+2},
\end{align}
which (if $s>-3$) evaluates to \begin{align}
     V_{PL}(a) =2^{s+1} \pi^{-1/2} \sigma_r^s \Gamma \left( \frac{s+3}{2} \right) \null_1 F_1 \left( -\frac{s}{2} ; \frac{3}{2} ; -\frac{a^2}{4\sigma^2} \right), \label{powerlaw}
\end{align}
where $_1 F_1 (a;b;z)$ is the Kummer confluent hypergeometric function.

\begin{figure}
    \centering
    \includegraphics[width=.75\textwidth]{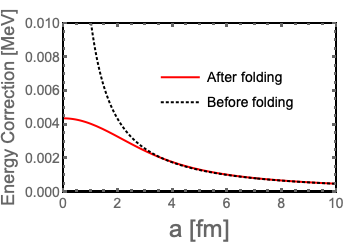}
    \caption{The correction to the Coulomb energy of two protons separated by a distance $a$ due to electron-positron pair production before folding (black, dotted), and after folding (red). We use $\sigma_r = 1.15\,\text{fm}$.}
    \label{fig:correction}
\end{figure}

\begin{figure}
    \centering
    \includegraphics[width=.75\textwidth]{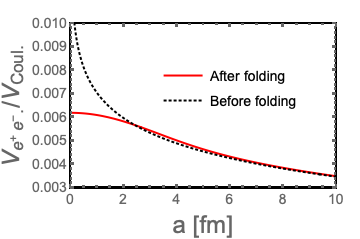}
    \caption{The fractional change of the Coulomb energy of two protons separated by a distance $a$ before folding (black, dotted), and after folding (red). (Note the vertical scale.) After folding, the correction is never more than about 0.6\%.}
    \label{fig:ratio}
\end{figure}

The correction to the Coulomb energy due to vacuum polarization (Equation \eqref{eq:uehlingpot}) contains two terms: one is proportional to the Coulomb energy, and the other is proportional to $\ln(r)/r$. We can fold the second term into the Gaussian by using the following identity
\begin{equation}
    \frac{\ln{r}}{r} = \pdv{r^s}{s}\Big|_{s=-1}.
\end{equation}
We conclude that the $\ln(r)/r$ potential folded into the Gaussian is related to the folded power law potential as
\begin{align}
    V_{log}(a) = \frac{\partial}{\partial s}  V_{PL}(a) \Big|_{s=-1}
\end{align}
Taking the derivative of equation \eqref{powerlaw} we find
\begin{align}
    V_{log}(a) = \frac{\ln{(2\sigma_r)} - \gamma/2}{a}\text{erf}\left(\frac{a}{2\sigma_r}\right) - \frac{1}{2\sqrt{\pi}\sigma_r} G^{(1)}\left(\frac{1}{2},\frac{3}{2};-\frac{a^2}{4\sigma_r^2} \right)\label{log}
\end{align}
where
\begin{align}
    G^{(1)}(a,b;z) = \frac{\partial}{\partial a}\null_1F_1(a,b;z)
\end{align}
From \eqref{eq:uehlingpot}, the folded vacuum polarization correction is
\begin{align}
    V_{e^+e^-}(a) = -\frac{2\alpha}{3\pi} \bigg[e^2 V_{log}(a) + \big(\gamma + 5/6 - \ln{\lambda_0} \big) V_c(a) \bigg]
\end{align}
Plugging in Eq. \eqref{powerlaw} gives our final result
\begin{multline}
    V_{e^+e^-}(a) = -\frac{2\alpha e^2}{3\pi} \Bigg[ \left(\gamma/2 + 5/6 + \ln{\frac{2\sigma_r}{\lambda_0}} \right) \frac{\text{erf}(a/2\sigma_r)}{a} \\ - \frac{1}{2\sqrt{\pi}\sigma_r} G^{(1)}\left(\frac{1}{2},\frac{3}{2}; -\frac{a^2}{4\sigma_r^2} \right) \Bigg]
\end{multline}
A plot of this potential is shown in Figure \ref{fig:correction}. The folding eliminates the $1/r$ singularity and approaches the point-particle expression for distances $\gtrsim 2 \sigma_r$. Figure \ref{fig:ratio} shows the correction as a fraction of the original Coulomb potential. In the point-particle approximation, this ratio diverges logarithmically at short distances, violating the assumption of a small electric field \cite{uehling}. Thus, Equation \eqref{eq:uehlingpot} is not valid at extremely short distances \cite{uehling}. Folding takes care of this issue; after folding, the correction is never more than about 0.6\% of the Coulomb energy.

\section{Conclusion}
This paper has been our first step in a theoretical research program to study dilepton production in strong fields. We have analyzed the vacuum polarization correction to the Coulomb energy in nuclear systems. Including a vacuum polarization term in the semi-empirical mass formula introduces a correction of about half a percent.
We have used the Bass potential as a simple phenomenological model to study the energies involved in collisions, and determined that the vacuum polarization energy is on the order of one percent of the total energy for heavy ions.
In future work, we will use Constrained Molecular Dynamics to get a more realistic idea of how vacuum polarization affects the dynamics.
We have been prepared for that task by deriving a formula for the vacuum polarization correction for protons with a Gaussian spatial distribution for use in the CoMD model. In the future, we also hope to calculate the $e^+e^-$ production rate in heavy ion collisions using the Schwinger mechanism \cite{schwinger,wong}.

\section*{Acknowledgements}
This work was supported by the United States
Department of Energy under Grant \# DE-FG03-93ER40773, NNSA DE-NA0003841 (CENTAUR), the National Natural Science Foundation of China No. 11 905 120 and the Fundamental Research Funds for the Central Universities (GK201903022).

\bibliography{mybibfile}

\end{document}